\documentclass{svproc}
\usepackage{xcolor}
\usepackage{graphicx}
\usepackage{multirow,multicol}
\usepackage{amsmath,amssymb,amsfonts}

\usepackage{sidecap}

\def\p{\partial}
\newcommand{\ud}{\mathop{}\!\mathrm{d}}
\newcommand\nc{\newcommand}
\nc\pad[2]{\frac{\p #1}{\p #2}} \nc\padd[2]{\frac{\p^2 #1}{\p
{#2}^2}} \nc\nd[2]{\frac{\ud #1}{\ud #2}} \nc\pat[2]{\frac{D #1}{D
#2}} \nc\ov{\overline} \nc\degree{^{\circ}} \nc\ord[1]{{\cal
O}(#1)} \nc\ra{\rightarrow} \nc\Ra{\Rightarrow} \nc\dint{{\mbox ~
d}}

\usepackage{soul}

\usepackage{url}

\begin{document}
\mainmatter              
\title{A mathematical model for contaminant adsorption in packed columns using nonlinear mass transfer}
\titlerunning{Contaminant adsorption using nonlinear mass transfer}  
%
\author{
F. Font\inst{1}, A. Estop\inst{2}, M. Calvo-Schwarzwalder\inst{3}, A. Valverde\inst{4}, T.G. Myers\inst{5}
}
\authorrunning{Font et al.} 
%
\tocauthor{}
\institute{
Department of Fluid Mechanics, Universitat Polit\`ecnica de Catalunya,\\ 
08019 Barcelona, Spain\\
\email{francesc.font@upc.edu}\\
\and
Universitat Aut\`onoma de Barcelona, Campus de Bellaterra, Edifici C,\\ 
08193 Bellaterra, Barcelona, Spain.\\
\email{ albert.estop@autonoma.cat}\\
\and
College of Interdisciplinary Studies, Zayed University,\\ 
P.O. Box 144534 Abu Dhabi, United Arab Emirates\\
\email{marc.schwarzwalder@zu.ac.ae}\\
\and
Department of Chemical Engineering, Universitat Polit\`ecnica de Catalunya,\\ 
08028 Barcelona, Spain\\
\email{abel.valverde@upc.edu}\\
\and
Centre de Recerca Matem\`atica, Campus de Bellaterra, Edifici C,\\ 
08193 Bellaterra, Barcelona, Spain.\\
\email{tmyers@crm.cat}\\
}

\maketitle              

\begin{abstract}
We propose a mathematical model for describing column adsorption processes where large amounts of contaminant are removed from a carrier gas. The large mass loss causes velocity and pressure to be variables which must then be determined along with the contaminant and gas concentrations and adsorbed quantity. The adsorption rate is described using a Langmuir sink, which is physically more sensible than the linear sink used in a previous study. The proposed model is solved numerically and using a traveling wave approach. The analytical solutions are consistent with previous models with fixed gas velocity and provide excellent agreement when validated against experimental data.

\keywords{Contaminant removal, Pollutant removal, Adsorption, Fluid dynamics, Langmuir equation}
\end{abstract}
\section{Introduction}
Column sorption is a process where a fluid, typically formed by a contaminant and a carrier fluid, is forced to flow through a column that is packed with a porous adsorbent material. The contaminant then attaches to the adsorbent until an equilibrium state is reached, where the concentration of contaminant measured at the outlet is equal to that of the inlet. When trace amounts are removed, such that the pressure and velocity of the carrier  are not affected the process is relatively well understood \cite{Valverde2023,Myers2023,Myers2020a,Aguareles2023}. When a significant quantity of the mixture is removed these previous models no longer hold. Myers and Font \cite{Myers2020b} proposed an extension  that accounts for larger amounts of contaminant, where the fluid velocity $u$ and pressure $p$ are unknown quantities to be determined along with the contaminant and carrier gas concentration profiles, respectively $c_{1}$ and $c_{2}$, and adsorbed amount per unit mass of adsorbent $q$. In \cite{Myers2020b} a standard linear kinetic model was employed to describe the mass removal however this neglects desorption and does not explicitly account for the presence of contaminant with the result that, theoretically, adsorption may occur without any contaminant to be adsorbed. The Langmuir model \cite{Langmuir1918} does not have these limitations,  which is why we propose extending the  model of \cite{Myers2020b} using this kinetic model.

\section{Mathematical formulation}
We consider a column of length $L$ that is packed with a porous material. A gas mixture with fixed contaminant and carrier gas volume fractions $\phi_1$ and $\phi_2$ is introduced at the inlet at a constant velocity $u_{in}$ and is adsorbed by the adsorbent as it flows towards the outlet, where the concentration of contaminant is measured. We assume that the pressure at the inlet is higher than the pressure at the outlet, $p_{in}=p_a+\Delta p$ and $p_{out}=p_a$, where $p_a$ is the ambient pressure. To maintain a constant mass flux, the pressure drop must be a function of time, we write $\Delta p=P(t)$. Using the ideal gas law, for a given volume fraction $\phi_i$, the concentration, $c$, of species $i$ just outside the inlet is
\begin{align}\begin{split}\label{InletConcentrations}
    c_{i}(0^-,t)=&\frac{\phi_ip_{in}(t)}{R_gT}=\frac{\phi_i}{R_gT}(p_a+P(t))\\
    =&\frac{\phi_ip_a}{R_gT}\left(1+\frac{P(t)}{p_a}\right)=c_{i0}\left(1+\frac{P(t)}{p_a}\right)\, ,
\end{split}\end{align}
where $T$ the temperature and $R_g$ is the universal gas constant. If the pressure drop is small ($P\ll p_a$) then the inlet concentrations can be considered approximately constant.

Following the model proposed by Myers and Font \cite{Myers2020b}, we consider all the quantities averaged over the column cross-section and therefore consider only one spatial variable $x$, representing the distance from the column inlet. Assuming that we are in a Darcy flow regime (this is shown in \cite{Myers2020b}), the dimensional model consists of the following set of equations:
\begin{subequations}\begin{align}
\pad{ c_{1}}{t}  &+ \pad{}{x}(u c_{1}  ) = D  \padd{ c_{1} }{x}  -  \rho_{q}   \pad{q}{t}   ~ ,\label{c1eq} \\
\pad{c_{2} }{t}  &+ \pad{}{x}(u c_{2}) = D  \padd{ c_{2}}{x} \, ,\\
p  & =    R_g T  ( c_{1} +   c_{2}) \, ,\quad
-\pad {p}{ x} = \frac{\mu}{\kappa}u
 \, , \quad
 \pad{q}{t}=\dot{Q}\, ,
\end{align}\end{subequations}
where $D$ is the mass diffusivity, $\rho_{q}$ is an effective density related to the porosity and density of the adsorbent, $\mu$ is the dynamic viscosity, and $\kappa$ is the permeability.
The expression of $\dot{Q}$ depends on the underlying mass transfer model. The Langmuir model  is
\begin{equation}
     \dot{Q}(q,c_{1})=k_{ad}c_{1}(q_{m}-q)-k_{de}q\, ,
\end{equation}
where $k_{ad}$ and $k_{de}$ are the adsorption and desorption rates and $q_m$ represents the maximum amount that may be adsorbed. Equilibrium is reached when both processes balance and thus contaminant concentration does not change with respect to its inlet value. Solving $\dot{Q}=0$ to find the equilibrium amount $q_e$ yields the Langmuir isotherm
\begin{equation}\label{Langiso}
    q_e = \frac{q_mK_Lc_{10}}{K_Lc_{10}+1}\, ,
\end{equation}
where $K_L=k_{ad}/k_{de}$ is referred to as the Langmuir equilibrium constant. 

The boundary conditions for the concentrations are
\begin{align}\label{bcinlet}
    u_{in}c_{i}(0^-,t)    &=  \left. \left(uc_{i} - D \pad{c_{i}}{x}\right)\right|_{x=0^+} \, ,\qquad \left.\pad{c_{i}}{x}\right|_{x=L}=0\, \qquad (i=1,2)\, .
\end{align}
Additionally, for the pressure we have
\begin{equation}
    p(0,t)=p_{in}(t)\,\qquad p(L,t)=p_a\, .
\end{equation}
Finally, the initial conditions are 
\begin{equation}
    c_1(x,0)=q(x,0)=0\, ,\qquad p(x,0)=p_0(x)\, ,\qquad c_2(x,0)=\frac{p_{0}(x)}{R_gT}\, ,
\end{equation}
where $p_0(x)=p_a+(1-x/L)P(0)$. Following Myers and Font \cite{Myers2020b}, we scale the variables with the following:
\begin{align}\begin{split}
    c_i&\sim c_{i0}\, ,\quad q\sim q_m\, ,\quad u\sim u_{in}\, , \quad 
    p-p_a\sim\mathcal{P}\, ,\quad x\sim\mathcal{L}\, ,\quad t\sim\tau\, ,
\end{split}\end{align}
where 
\begin{equation}
    \mathcal{P}=\frac{\mu\mathcal{L}u_{in}}{\kappa}\, ,\qquad \mathcal{L}=\frac{u_{in}\tau c_{10}}{\rho_qq_m}\, ,\qquad \tau=\frac{1}{c_{10}k_{ad}}\, .
\end{equation} 
Upon indicating the scaled quantities with a hat notation and following the nomenclature of \cite{Myers2020b}, the non-dimensional formulation of the problem reads
\begin{subequations}\begin{align}
\delta_1\pad{ \hat c_{1}}{t}  &+ \pad{}{\hat x}(\hat u \hat c_{1}  ) = \delta_2  \padd{ \hat c_{1} }{\hat x}  - \pad{\hat q}{\hat t}\, , \qquad
\delta_1\pad{\hat c_{2} }{\hat t}  + \pad{}{\hat x}(\hat u \hat c_{2}) = \delta_2  \padd{ \hat c_{2}}{\hat x} \, ,\label{nondimEq1}\\
1+\delta_3\hat p  & = \delta_{45}\hat c_{1} +  \delta_4\hat c_{2} \, ,\quad
-\pad {\hat p}{\hat x} = \hat u\, , \quad
 \pad{\hat q}{\hat t}=\hat c_1(1-\hat q)-\delta_6\hat q\, ,\label{nondimEq2}
\end{align}\end{subequations}
where
\begin{align}\begin{split}
    &\delta_1=\frac{\mathcal{L}}{\mathcal{U}\tau}\, ,\qquad\,\,\,\,\,\, \delta_2=\frac{D}{\mathcal{UL}}\, ,\quad \delta_3=\frac{\mathcal{P}}{p_a}\, ,\\  
    \delta_4=\phi_2\, ,&\quad \delta_5=\frac{\phi_1}{\phi_2}\, , \quad \delta_{45}=\delta_{4}\delta_{5}=\phi_1\, , \quad \delta_6=\frac{1}{K_Lc_{10}}\, .
\end{split}\end{align}
The non-dimensional boundary conditions are
\begin{subequations}\begin{align}\label{bcinlet}
    &1+\delta_3\hat P(\hat t)    =  \left. \left(\hat u\hat c_{j} - \delta_2 \pad{\hat c_{j}}{\hat x}\right)\right|_{\hat x=0} \, ,\qquad \left.\pad{\hat c_{j}}{\hat x}\right|_{\hat x=\hat L}=0\, \qquad (j=1,2)\, ,\\
    &\hat p(0,\hat t) = \hat P(\hat t)\, ,\qquad \hat p(\hat L,\hat t)=0\, ,\qquad \hat p(\hat x,0) =\left(\hat{L}-\hat x\right)\hat P(0)\, ,\\
    &\hat c_1(\hat x,0)=\hat q(\hat x,0)=0\, ,\qquad \delta_4 \hat c_2(\hat x,0)=1+\delta_3\hat p(\hat x,0)\, ,
\end{align}
\end{subequations}
with $\hat L=L/\mathcal{L}$ and $\hat P=P/\mathcal{P}$.

The parameters $\delta_1$ and $\delta_2$ are the Damk\"ohler and inverse P\`eclet numbers, $\delta_3$ is the relative pressure, $\delta_4$, is the volume fraction of carrier gas in the mixture before entering the column, $\delta_5$ is the ratio of the volume fractions, and $\delta_6$ relates adsorption and desorption. Following the discussion of \cite{Myers2020b}, we assume $\delta_1,\delta_2,\delta_3\ll1$ so that the corresponding terms are negligible to leading order.

To retrieve the fixed velocity model provided in \cite{Myers2023}, we will have to take the limits $\phi_1=\delta_{45}\to0$ and $\phi_2=\delta_4\to1$.

\section{Solution Methods}

\subsection{Traveling wave solution}
\label{tw_section}
These kinds of models may typically be solved using a traveling wave approach \cite{Valverde2023,Myers2023,Myers2020a,Aguareles2023,Myers2020b}. Upon neglecting terms of order $\delta_1,\delta_2,\delta_3$, and writing $\delta_4$ and $\delta_5$ in terms of the contaminant void fraction $\phi_1$, the problem reduces to
\begin{subequations}\begin{align}
&\pad{}{\hat x}(\hat u \hat c_{1}  ) = - \pad{\hat q}{\hat t}\, , \qquad
\pad{}{\hat x}(\hat u \hat c_{2}) =0 \, ,\label{red:nondimEq1}\\
1 = \phi_1\hat c_{1} +&  (1-\phi_1)\hat c_{2} \, ,\quad
-\pad {\hat p}{\hat x} = \hat u\, , \quad
 \pad{\hat q}{\hat t}=\hat c_1(1-\hat q)-\delta_6\hat q\, .\label{red:nondimEq2}
\end{align}\end{subequations}
From the reduced equations we find
\begin{equation}\label{TWreduction}
    \hat c_2 = \frac{1-\phi_1\hat c_1}{1-\phi_1}\, ,\qquad \hat u=\frac{1-\phi_1}{1-\phi_1\hat c_1}\, ,
\end{equation}
where the first comes from rearranging Eq. (\ref{red:nondimEq2}.i) and the second from integrating Eq. (\ref{red:nondimEq1}.ii) and applying the boundary conditions at $\hat x=0$. The problem then reduces to solving
\begin{equation}
    \pad{}{\hat x}\left(\frac{(1-\phi_1)\hat c_{1}}{1-\phi_1\hat c_1}\right) = - \pad{\hat q}{\hat t}\, , \qquad \pad{\hat q}{\hat t}=\hat c_1(1-\hat q)-\delta_6\hat q\, .
\end{equation}
Now we introduce the traveling wave coordinate $\eta=\hat x-\hat L - \hat v(\hat t-\hat t_{1/2})$, where $\hat v$ is constant and $\hat t_{1/2}$ is the (dimensionless) time when the outlet concentration is half of the inlet value, $c_1=c_{10}/2$. Upon defining $(1-\phi_1)\hat c_1/(1-\phi_1\hat c_1) = F(\eta)$ and $\hat q=G(\eta)$ and denoting derivatives with respect to $\eta$ with primes, the problem becomes
\begin{equation}
    F'= \hat v G'\, , \qquad -\hat vG'=\frac{F(1-G)}{1-\phi_1(1-F)}-\delta_6G\, ,
\end{equation}
where we impose $F\to1$, (i.e. $\hat c_1\to1$) $G\to1/(1+\delta_6)=G_e$ as $\eta\to-\infty$ (equilibrium far behind the wavefront) and $F,G\to0$ as $\eta\to\infty$ (no free or adsorbed contaminant far ahead of the wavefront). Note, $\hat c_1=1/2$ at $\eta=0$ yields the additional condition $F(0)=(1-\phi_1)/(2-\phi_1)=F_{1/2}$. Using these conditions, we can solve the problem and find
\begin{equation}
    G=\frac{F}{\hat v}\, ,\qquad \hat v=\frac{1}{G_e}=1+\delta_6\, ,
\end{equation}
and $F$ is given implicitly by the equation
\begin{equation}\label{TWsoln}
    -\frac{\eta}{1+\delta_6}
    = \frac{1-\phi_1}{1+\phi_1\delta_6}\ln\left|\frac{ F}{ F_{1/2}}\right|
    -\frac{1}{1+\phi_1\delta_6}\ln\left|\frac{1-F}{1- F_{1/2}}\right|\, .
\end{equation}
In terms of $\hat c_1$ this yields
\begin{equation}\label{TWsoln_c}
    -\frac{\eta}{1+\delta_6}
    =\frac{1}{1+\phi_1\delta_6}\ln\left|\frac{\hat c_1}{1-\hat c_1}\right|-\frac{\phi_1}{1+\phi_1\delta_6}\ln\left|\frac{(2-\phi_1)\hat c_1}{1-\phi_1\hat c_1}\right| \, .
\end{equation}
This method determines $\hat c_1$, $\hat c_2$, $\hat q$, and $\hat u$. Their profiles are shown in Fig. \ref{fig:twsoln} and later discussed in section \S \ref{resdis}. The form in which Eq. \eqref{TWsoln_c} is presented allows the fixed velocity model to be easily retrieved by taking the limit $\phi_1\to0$. This limit is also discussed in \S\ref{resdis}

At this point, the pressure is the only quantity that has not been determined. We find it by integrating Eq. (\ref{red:nondimEq2}.ii) and applying $\hat p=0$ at $\hat x=\hat L$. The pressure drop $\hat P$ may then be found by applying the boundary condition at the inlet. We find
\begin{equation}
\hat p=\int_{\hat x}^{\hat L}\frac{1-\phi_1}{1-\phi_1\hat c_1(\xi,\hat t)}\ud\xi\, ,\qquad \hat P(\hat t)=\int_{0}^{\hat L}\frac{1-\phi_1}{1-\phi_1\hat c_1(\xi,\hat t)}\ud\xi\, . 
\end{equation} 

Upon substituting $\eta(\hat L,t)=\hat v(\hat t-\hat t_{1/2})$, we find the implicit breakthrough model
\begin{equation}\label{TW:breakthrough}
    \hat t = \hat t_{1/2} + f\left(\hat c_1(\hat L,\hat t)\right)\Rightarrow t = t_{1/2} + \frac{1}{k_{ad}c_{10}}f\left(\frac{c_1(L,t)}{c_{10}}\right)\, ,
\end{equation}
where $f(\cdot)$ is defined as the right-hand side of Eq. \eqref{TWsoln_c}.

\subsection{Numerical solution}


We expect $\delta_3=\ord{10^{-2}}$, hence setting $\delta_3=0$ suggests errors of the order of 1\%. Upon doing this, the model unknowns reduce to $\hat c_1$, $\hat q$ and $\hat u$, while $\hat c_2$ and $\hat p$ become passive variables that can be obtain once $\hat c_1$, $\hat q$ and $\hat u$ are known. Then, we can easily adapt the numerical method developed in \cite{Myers2020b}, which is based on solving for $\hat c_1$, $\hat q$ and $\hat u$ using second-order central finite differences in space and explicit Euler in time, while $\hat c_2$ and $\hat p$ are constructed {\it a posteriori}. Note that this approach retains the terms with $\delta_1$ and $\delta_2$ that were neglected in the analytical approach from section \S \ref{tw_section}. Thus, the numerical solution can be used to validate the accuracy of the travelling wave solution. In order to have a consistent comparison with the travelling wave solution, the approximation $u\rightarrow 1$ as $x\rightarrow 0$ is used in the boundary condition (\ref{bcinlet}.i).

\subsection{Results and discussion }
\label{resdis}

\begin{figure}
    \centering
    \includegraphics[width=.49\textwidth]{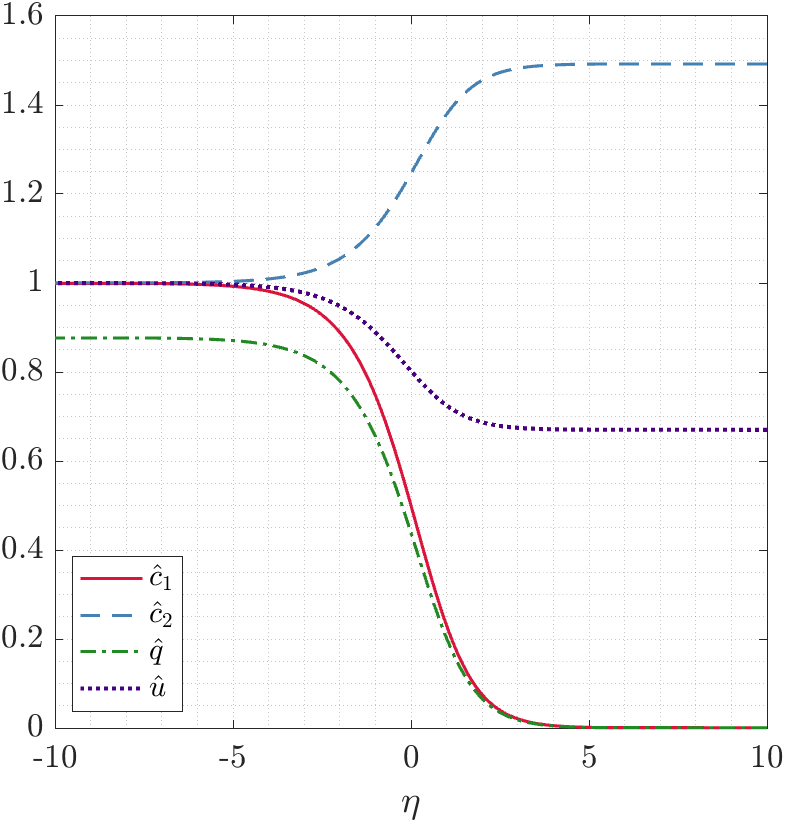}
    \hfill     
    \includegraphics[width=.49\textwidth]{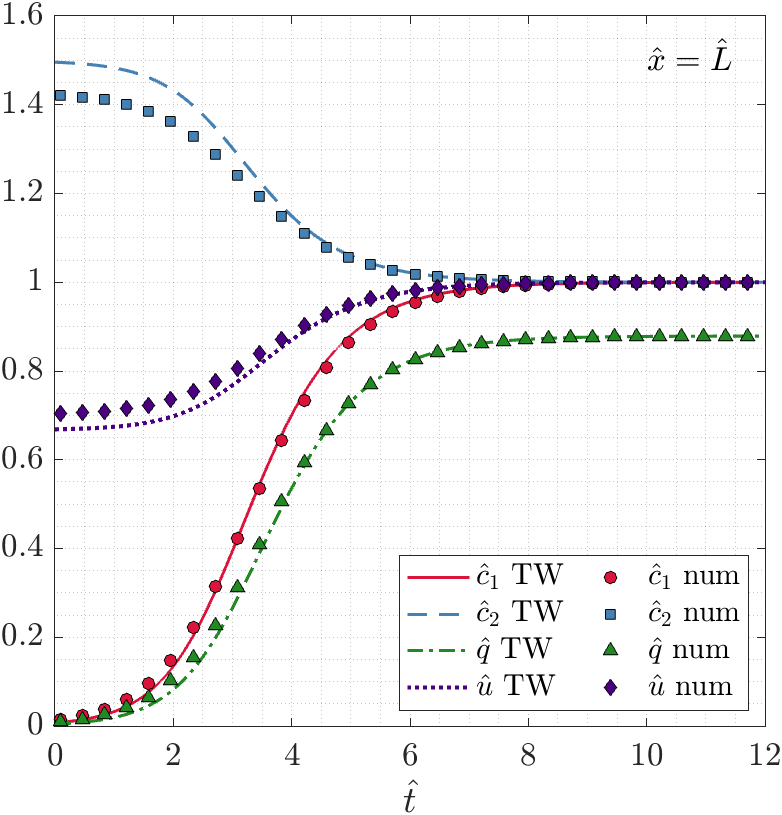}\\
    \caption{On the left, profiles of the concentrations, adsorbed amount and velocity according to the traveling wave solutions for $\phi_1 = 0.33$ and $\delta_6=0.14$, which is consistent with the experimental work used in \S\ref{ssec:carbon}. On the right, comparison of breakthrough curves computed numerically and via Eq. \eqref{TW:breakthrough}, using the parameter values found in Table \ref{tab:optimization} and the non-dimensional parameters provided in \S\ref{ssec:carbon}. The dimensionless half-time is $\hat t_{1/2}=3.32$. }
    \label{fig:twsoln}
\end{figure}

The left panel from Fig. \ref{fig:twsoln} shows the variation of the concentration of  contaminant and  carrier gas through the column, along with the adsorbed fraction and the velocity profile of the gas. As imposed by the boundary conditions the concentration of contaminant is close to one for large negative $\eta$ and decreases to zero for large positive $\eta$. Similarly, the adsorbed fraction tends to its constant equilibrium value near the inlet, $\hat q \approx 0.88$, and to zero near the wave front where the adsorbent material is still clean of contaminant (i.e, $\hat q = 0$). In the region of the column where the adsorbent is full of contaminant the gas velocity is highest and it decreases towards the clean region of the column, indicating that adsorption slows down the gas transport. Finally, the right panel from Fig. \ref{fig:twsoln} shows the results at the outlet, comparing the numerical and travelling wave solutions. The key quantities $\hat{c}_1$, $\hat{q}$ show very good agreement, except for small times where the numerical solution is expected to be more accurate due to the inclusion of the terms with $\delta_1$ and $\delta_2$ which were neglected in the derivation of the travelling wave solution. 

\begin{figure}
    \centering
    \includegraphics[width=.49\textwidth]{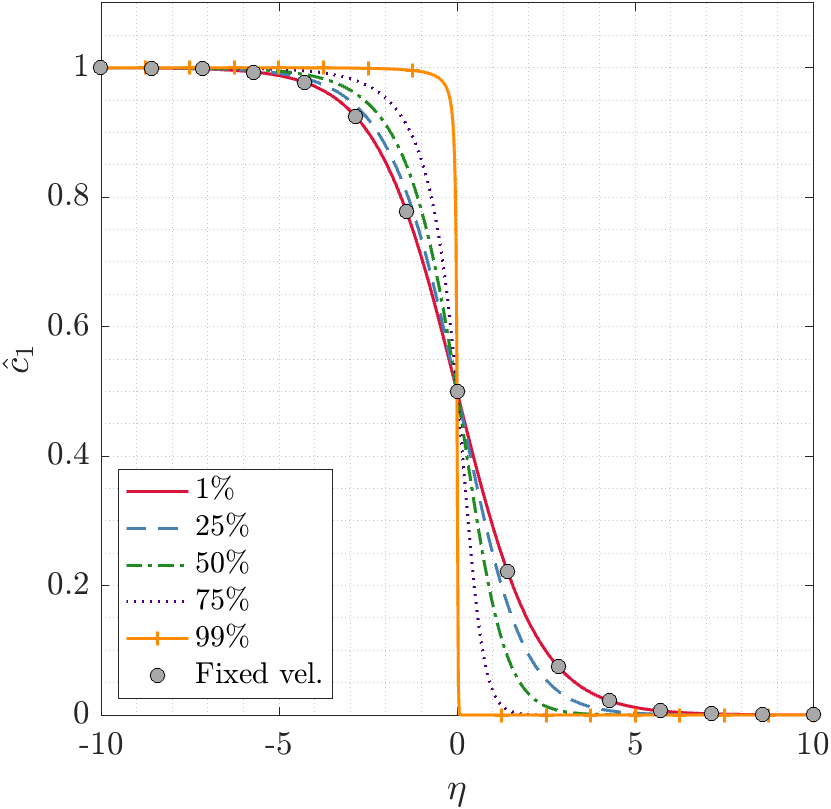}
    \hfill     
    \includegraphics[width=.49\textwidth]{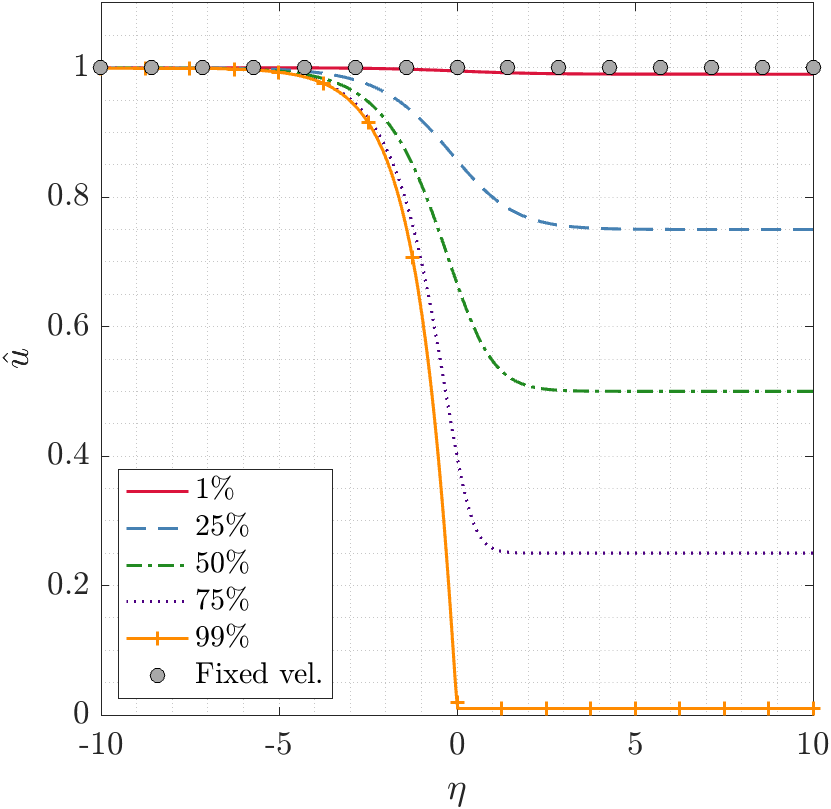}
    \caption{Evolution of the contaminant concentration and gas velocity in terms of the traveling wave co-ordinate for different volume fractions $\phi_1$ and $\delta_6=0.14$. The circles represent the solution of the fixed-velocity model described in \cite{Myers2023}.}
    \label{fig:volumefraction}
\end{figure}

In Fig. \ref{fig:volumefraction} we show the contaminant concentration and velocity profiles for a range of volume fractions $\phi_1$. For 1\% of contaminant, the profiles agree with fixed-velocity solution from \cite{Myers2023}. Actually, the previous model can be obtained analytically from the present model by taking the limit $\phi_1\to0$ in Eq. \eqref{TWsoln_c}, which yields $\hat c_1=1/\left(1+e^{\eta/\hat v}\right)$, leading to the fixed-velocity dimensional breakthrough model of \cite{Myers2023},
\begin{equation}\label{TW:breakthrough0}
    c_1(L,t) = \frac{c_{10}}{1+e^{k_{ad}c_{10}(t_{1/2}-t)}}\, .
\end{equation}

For small values of $\phi_1$, from Eq. (\ref{TWreduction}.ii) we find 
\begin{equation}
    \hat u=(1-\phi_1)\left(1+\phi_1\hat c_1+\ord{\phi_1^2}\right)=1-\phi_1(1-\hat c_1)+\ord{\phi_1^2}\, ,
\end{equation}
and thus the constant velocity is recovered at leading order. Conversely, for large values of $\phi_1$ we find that both the velocity and contaminant concentration tend to constant values at either side of the wavefront, indicating a small boundary layer around the origin $\eta=0$ for $\phi_1\to0$. We can observe how the maximum velocity variation $\Delta\hat u$ is directly related to the volume fraction occupied by the contaminant:
\begin{equation}
    \Delta\hat u=\hat u(-\infty)-\hat u(\infty) = \phi_1\, ,
\end{equation}
which provides a justification for using the variable-velocity model when the volume fraction is significant.

\section{Application to CO$_\text{2}$ capture}\label{ssec:carbon}

We now validate the variable-velocity model against experimental data. For this we use the experimental data of Monazam \emph{et al.} \cite{Monazam2013}, who capture CO$_2$ using polyethylenimine (PEI) modified silica from a gas mixture formed by the contaminant and N$_2$ at 90 $^\circ$C. The relevant operating conditions are summarized in Table \ref{tab:Monproperties}.

\begin{table}
    \centering
    \caption{Operating conditions and fluid parameter values from \cite{Monazam2013}. Isotherm parameters calculated by interpolating the values at 80$^\circ$C and 100$^\circ$C provided in \cite{Monazam2013b}. Leading order inlet concentrations are computed using Eq. \eqref{InletConcentrations}.}
    \label{tab:Monproperties}
    \begin{tabular}{lccc}\hline
        \textbf{Property} & \textbf{Symbol} & \textbf{Units} & \textbf{Value} \\ \hline
        Column length & $L$ & m & $0.254$ \\
        Effective density & $\rho_q$ & kg/m$^3$ & $2100$ \\
        Dispersion coef. & $D$ & m$^2$/s & $4.8\times10^{-5}$ \\
        Inlet fluid vel. & $u_0$ & m/s & $0.0944$ \\
        Ambient pressure & $p_a$ & Pa (Atm) & $101325$ ($1$) \\
        Temperature & $T$ & K & $363.15$ \\
        Gas viscosity & $\mu$ & Pa$\cdot$s & $2.07\times10^{-5}$ \\
        Permeability & $\kappa$ & m$^2$ & $2.76\times10^{-11}$ \\
        CO$_2$/N$_2$ inlet conc. & $c_{10}$/$c_{20}$ & mol/m$^3$ & $11.18$/$22.38$\\
        CO$_2$/N$_2$ vol fraction & $\phi_1$/$\phi_2$ & - & $0.333$/$0.667$ \\
        Final adsorbed fraction & $q_{e}$ & mol/kg & $2.21$ \\
        Max. adsorbed fraction & $q_{m}$ & mol/kg & $2.52$ \\
        Equilibrium constant & $K$ & m$^{3}$/mol & 0.65 \\
        \hline
    \end{tabular}
\end{table}

\begin{figure}
    \centering
    \includegraphics[width=.98\textwidth]{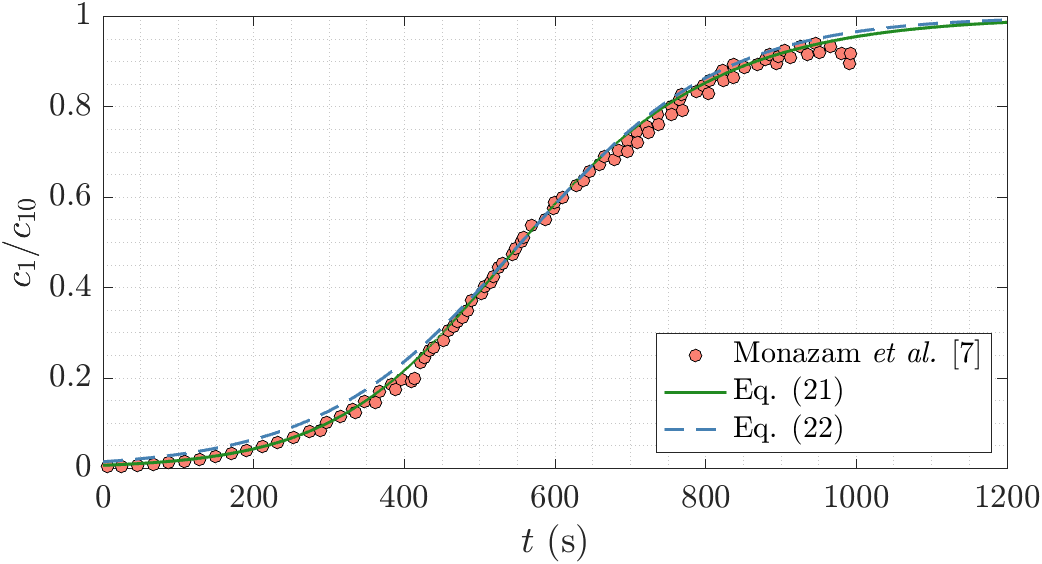}
    \caption{
    Fitting of the model to the breakthrough data reported by Monazam \emph{et al.} \cite{Monazam2013}, using fixed and variable velocity models, at 90$^\circ$C, 30 standard L/min and 33.3\% of CO$_2$ at the inlet flow. 
    }
    \label{fig:results}
\end{figure}

In Fig. \ref{fig:results} we have fitted the value of $k_{ad}$ for both the new and previous models of Eqs. \eqref{TW:breakthrough} and \eqref{TW:breakthrough0}, with the optimization results shown in Table \ref{tab:optimization}. By comparing the error values, we can observe that the present model is more accurate than the fixed-velocity model. The accuracy is especially high for small times, indicating that the reduction in velocity during the process (shown in Fig. \ref{fig:volumefraction}b) is  significant. If the ratio of concentrations $\phi_1/\phi_2$ were to increase we anticipate that the fitting of equation \eqref{TW:breakthrough0} would become less accurate while the present model would remain accurate. The non-dimensional parameters are
 $\delta_1=2.11\times 10^{-3}$, $\delta_2=2.05\times 10^{-2}$, $\delta_3=1.48\times 10^{-2}$, $\delta_4=0.667$, $\delta_5=0.499$, and $\delta_6=0.138$. Their orders of magnitude clearly justify the traveling wave approach.

\begin{table}
    \centering
    \caption{Comparison between the fitting to Monazam \emph{et al.} \cite{Monazam2013} breakthrough data of the variable and constant velocity models for 90$^\circ$C.}
    \label{tab:optimization}
    \begin{tabular}{ccc}
    \hline
      \textbf{Parameter} & \textbf{Value ($u$ var.)} & \textbf{Value ($u$ cons.)}  \\ \hline
      $k_{ad}$ (m$^3$mol$^{-1}$s$^{-1}$) & 5.352$\times 10^{-4}$ & 6.753$\times 10^{-4}$ \\ 
      SSE & 0.0245 & 0.0611\\ 
      R-squared & 0.9976 & 0.9941 \\ \hline
    \end{tabular}
\end{table}

    

\section{Conclusion}
We have presented an extension to the basic Langmuir adsorption model that accounts for the removal of large amounts of contaminant, such that the fluid velocity is variable. The solutions obtained using the traveling wave approach are consistent with the fixed-velocity model in \cite{Myers2023}. The model is restricted to small pressure drops, which yield approximately constant inlet concentrations for fixed volume fractions. In practise, this does not represent a limitation, since the relevant experiments show very slow velocities and hence small pressure drops.

The experimental section provides evidence of the good agreement between the new model and the breakthrough data from a column adsorption  process for the capture of CO$_2$. In particular, the agreement is better than for the previous fixed-velocity approach. The model presented is consistent with the results obtained by fitting the isotherm. The results show how the constant velocity model loses accuracy when trying to fit the breakthrough curve obtained with a high percentage of CO$_2$ in the inlet flow is low.

Further work will include validation with larger sets of experimental data with different volume fractions. The present model only accounts for physical adsorption, hence it will be generalized to account for chemical adsorption as well. This may be done using the Sips kinetic model, which has been recently used under the assumption of constant flow \cite{Aguareles2023}. For developing industrial-scale technologies, intra-particle diffusion may also be a relevant factor to introduce in the future.

\section*{Acknowledgements}
This publication is part of the research projects PID2020-115023RB-I00 and TED2021-131455A-I00 (funding T. Myers and F. Font) financed by
MCIN/AEI/ 10.13039/501100011033/, by “ERDF A way of making Europe” and by “European Union NextGenerationEU/PRTR”. A. Valverde acknowledges support from the Margarita Salas UPC postdoctoral grants funded by the Spanish Ministry of Universities with European Union funds - NextGenerationEU. T. Myers acknowledges the CERCA Programme of the Generalitat de Catalunya. The work was also supported by the Spanish State Research Agency, through the Severo Ochoa and Maria de Maeztu Program for Centres and Units of Excellence in R\&D (CEX2020-001084-M). F. Font is a Serra-Hunter fellow from the Serra-Hunter Programme of the Generalitat de Catalunya.

%

\end{document}